\documentclass[sigconf,nonacm]{acmart}

\usepackage{pvldb}

\renewcommand\vldbdoi{XX.XX/XXX.XX}
\renewcommand\vldbpages{XXX-XXX}
\renewcommand\vldbavailabilityurl{}
\usepackage{booktabs}
\usepackage{multirow}
\usepackage{graphicx}
\usepackage{array}
\usepackage{tabularx}
\usepackage{xspace}
\usepackage{algorithm}
\usepackage{algpseudocode}

\algrenewcommand\algorithmicrequire{\textbf{Input}}
\algrenewcommand\algorithmicensure{\textbf{Output}}

\usepackage[dvipsnames,table]{xcolor}
\definecolor{LightCyan}{RGB}{232,241,255}
\definecolor{WhiteColr}{RGB}{255,255,255}
\definecolor{BestGreen}{RGB}{0,140,64}

\newcommand{\dltup}[1]{\textcolor{OliveGreen}{#1}}   
\newcommand{\dltdn}[1]{\textcolor{BrickRed}{#1}}     
\newcommand{\mvup}[2]{\dltup{$#1\,(#2)$}}
\newcommand{\mvdn}[2]{\dltdn{$#1\,(#2)$}}
\newcommand{\mvupb}[2]{\dltup{$\mathbf{#1}\,(#2)$}}

\newcommand{\na}{--}
\begin{document}

\title{SmartANN: Object Causal Modeling Boosts Approximate Nearest Neighbor Diagnosis and Auto-Design}

\author{Yutong Zhou}
\affiliation{%
      \institution{University of Chinese Academy of Sciences (ICT,CAS)}
  \city{Beijing}
  \country{China}
}
\email{zhouyutong24s@ict.ac.cn}

\author{Guoxin Kang}
\authornote{Corresponding author.}
\affiliation{%
  \institution{Institute of Computing Technology, Chinese Academy of Sciences}
  \city{Beijing}
  \country{China}
}
\email{kangguoxin@ict.ac.cn}

\author{Lei Wang}
\affiliation{%
  \institution{Institute of Computing Technology, Chinese Academy of Sciences}
  \city{Beijing}
  \country{China}
}
\email{wanglei\_2011@ict.ac.cn}

\author{Xueya Zhang}
\affiliation{%
  \institution{University of Chinese Academy of Sciences (ICT,CAS)}
  \city{Beijing}
  \country{China}
}
\email{zhangxueya21@mails.ucas.ac.cn}

\author{Qinwei Yang}
\affiliation{%
  \institution{The University of Hong Kong}
  \city{Hong Kong}
  \country{China}
}
\email{yangqinwei2003@connect.hku.hk}

\author{Jianfeng Zhan}
\affiliation{%
  \institution{Institute of Computing Technology, Chinese Academy of Sciences}
  \city{Beijing}
  \country{China}
}
\email{zhanjianfeng@ict.ac.cn}

\begin{abstract}
Approximate Nearest Neighbor (ANN) algorithms achieve high efficiency through multiple interdependent phases across index construction and query execution.
However, this tight coupling allows performance loss from an upstream phase to propagate to downstream phases, affecting both their execution behavior and measurable outputs. 
Existing component-level works analyze mainly isolate and compare algorithmic design choices, while end-to-end benchmarks only aggregate performance metrics. 
Neither traces performance-loss propagation across
dependent phases, making it difficult to attribute root causes or automatically redesign.

To address this challenge, we present SmartANN, a framework built on the object causal model (OCM) for ANN bottleneck attribution and automated redesign. SmartANN represents the ANN workflow as eight ordered, replaceable objects and diagnoses them through a sequential diagnose-and-replace loop. At each iteration, it identifies the first object that deviates from the expected behavior or output, which we call a test oracle, as a bottleneck. As the upstream bottleneck object will obscure the downstream bottlenecks, SmartANN replaces the former with the test oracle if it exists,  or an implementation with a better outcome otherwise, and then diagnoses the downstream objects. 
Based on the resulting bottleneck set and failure causes, SmartANN automatically selects and composes compatible actions from a pluggable action library to cover all diagnosed bottlenecks and generate an optimized end-to-end ANN design. We instantiate SmartANN for IVF-PQ and HNSW, representing widely used partition-and-quantization and graph-based ANN families, respectively.

Extensive experiments on eight real-world datasets demonstrate that SmartANN improves Recall by 0.24--74.20\%, while it increases QPS by 28.8--256.5\% at comparable Recall, with low diagnosis and auto-design overhead. The code is available at \url{https://github.com/zhouyutong20/SmartANN}.

\end{abstract}

\maketitle

\vldbtopmatter

\section{Introduction}

ANN algorithms are essential for high-dimensional vector retrieval\\~\cite{surveyofvb,Han2023ACS,annonhigh}. They are widely used in recommendation systems~\cite{rec16,recKDD17}, information retrieval~\cite{informationretrieval}, and retrieval-augmented generation for large language models~\cite{rag1,rag2}.
However, the tight inter-phase dependencies in ANN algorithms cause \textit{cascading performance losses}, whereby upstream performance losses propagate downstream, complicating diagnosis.  
Existing component-level analyses mainly isolate and compare predefined design choices in graph-based algorithms~\cite{wang-survey-vldb2021}. Such comparisons reveal the effects of individual designs but do not trace how a performance loss propagates across dependent phases.
End-to-end benchmarks such as ANN-Benchmarks~\cite{annbench}, Big-ANN-Benchmarks~\cite{bigannbench}, and VectorDBBench~\cite{vecdbbench} report aggregate Recall, QPS, construction time, and memory footprint. They reveal whether an ANN algorithm performs poorly but not why it performs poorly.
Consequently, neither diagnosis can identify which internal phases cause the performance bottleneck under the current dataset or determine which existing ANN optimization can address it.

The lack of diagnosis also prevents automated ANN design.
Existing methods optimize different phases of ANN construction and
query execution. LAET \cite{laet} and DARTH \cite{darth} learn search
termination conditions. Neural LSH \cite{lsh2020iclr} and BLISS
\cite{blissKDD22} improve data partitioning. LTR-IVF
\cite{ltr-ivf} optimizes query routing. RPQ \cite{rpqICDE24}
improves vector quantization. ADSampling \cite{adsamplingSIGMOD23}
reduces distance-computation cost through adaptive sampling.
Because performance gains on one dataset frequently fail to generalize to another, users are forced into costly trial-and-error cycles to find the right optimization for their specific observed bottleneck.
\textit{Systematically diagnosing and automatically redesigning ANN pipelines} presents three core challenges. 

\textbf{\textit{(C1) Performance loss propagates through ANN execution.}} The output of an upstream phase becomes the input to downstream phases. An inefficient implementation in the upstream phases thus contaminates downstream inputs and confounds their diagnostic metrics. Consequently, independent phase inspection often fails because downstream bottlenecks are obscured by performance degradation cascading from upstream.

\textbf{\textit{(C2) Difficulty of Preventing the Propagation from the Upstream Phases and Exposing the Downstream Bottlenecks.}} To properly evaluate downstream execution, one must halt the upstream propagation of performance loss. This requires replacing the bottleneck phase with an implementation that can expose the bottlenecks in downstream phases. Because intermediate phases like data partitioning and graph construction lack theoretically exact outputs, so-called test oracles that can be computed directly, it is a major challenge to prevent the performance loss from propagating from the upstream phase to the downstream phase.

\textbf{\textit{(C3) Existing ANN optimizations are difficult to reuse for a specific executed phase.}} Existing ANN optimizations tightly couple modifications across multiple execution phases. This makes it exceedingly difficult to decompose them into modular, pluggable auto-design actions. Even when a specific bottleneck phase is diagnosed, users cannot easily extract and apply the corresponding repair.

To tackle the above challenges, we present SmartANN, a framework built on the object causal model (OCM) for ANN bottleneck attribution and automated redesign. SmartANN structures ANN index construction and query execution into eight ordered, replaceable objects, each defining its mechanism, diagnostic metrics, and available actions.
SmartANN diagnoses objects in execution order and records the first object that deviates from its expected behavior or output, which we call the test oracle, as a bottleneck. Because an upstream bottleneck may obscure downstream bottlenecks, SmartANN replaces its output with the test oracle, when available, or its implementation with a stronger replacement implementation otherwise, before continuing downstream diagnosis. The OCM replacement operator quantifies the downstream effect by comparing diagnostic metrics before and after the replacement. Repeating this \textit{diagnose-and-replace} loop yields the complete bottleneck-object set and failure causes.
Given this bottleneck set and causes, SmartANN automatically constructs candidate designs from individual optimization actions or compatible action compositions whose joint coverage spans the complete set. It instantiates and evaluates each candidate and selects the best-performing end-to-end design. Rather than proposing new optimization methods, SmartANN organizes existing ANN optimizations in a pluggable action library. We instantiate SmartANN for IVF-PQ and HNSW, representing widely used partition-and-quantization and graph-based ANN families, respectively.

\noindent\textbf{Contributions. }This paper makes the following contributions.

\noindent (1) We present an object causal model for ANN algorithms based on Evaluatology\cite{zhan2025evaluatology,evaluatology-2026-e}, and instantiate it for IVF-PQ and HNSW. We define a structured representation for each object, including mechanisms, actions, and diagnostic metrics, which makes internal performance loss measurable and analyzable at the object level.

\noindent (2) We propose object-level attribution through a sequential diagnosis-and-replace loop. In each iteration, SmartANN locates the first bottleneck object, replaces it with a test oracle or stronger replacement implementation, and uses the replacement operator to measure changes in downstream diagnostic metrics. Repeating this process identifies the complete bottleneck object set and failure causes.

\noindent (3) We build a pluggable library of 42 auto-design actions, with design knowledge encoded in action records and implementations standardized through object interfaces. Using the diagnosed bottleneck-object set and failure causes, SmartANN automatically generates and selects compatible action compositions to produce an optimized end-to-end ANN design.

\noindent (4) Extensive experiments on eight real-world datasets across IVF-PQ and HNSW demonstrate that bottleneck-object replacement effectively mitigates performance-loss propagation to downstream objects. SmartANN improves Recall by 0.24--74.20\% and increases QPS by 28.8--256.5\% at comparable Recall compared to the baseline. Compared with VDTuner, the state-of-the-art automatic performance tuning framework, SmartANN achieves a \(2.9\)--\(42.0\times\) speedup in end-to-end latency, demonstrating low diagnosis and auto-design overhead.

\section{Preliminaries}
\subsection{Problem Definition}
Given a dataset \(D=\{x_1,\ldots,x_n\}\subset\mathbb R^d\), a query set \(Q\subset\mathbb R^d\), a distance function \(\delta\), and an integer \(k\), an approximate nearest neighbor algorithm returns a result set \(\widehat N_k(q)\subseteq D\) of \(k\) vectors for each query \(q\in Q\). Let \(N_k(q)\) denote the exact \(k\) nearest neighbors of \(q\) in \(D\) according to \(\delta\). We measure search accuracy by

\begin{equation}
    \operatorname{Recall@k}
=
\frac{1}{|Q|}
\sum_{q\in Q}
\frac{|\widehat N_k(q)\cap N_k(q)|}{k}.
\end{equation}

We measure search efficiency by queries per second (QPS) under a fixed hardware environment. Each evaluated parameter configuration produces one Recall--QPS point. The non-dominated points form the Recall--QPS Pareto frontier of an ANN algorithm.

Given a baseline ANN algorithm, SmartANN diagnoses the internal execution phases responsible for its performance bottlenecks on \(D\) and \(Q\), and automatically generates redesigned ANN algorithms. For each evaluated configuration, the parameter values remain fixed, and the redesign aims to improve Recall, QPS, or both. 

\subsection{Object Causal Model}

The object causal model (OCM) models the causal effect propagation paths among different objects within a system~\cite{zhan2025evaluatology,evaluatology-2026-e}. The output of an upstream object can affect both the execution behavior and measurable outputs of downstream objects. OCM supports object-level replacement of mechanisms and actions to isolate these causal effects. SmartANN therefore adopts OCM for modeling the cause-and-effect relationship among the objects within the complex system, and uses the replace operator the measure the downstream effects of their replacements and guide automated ANN design.

\begin{figure*}[t]
    \centering
    \IfFileExists{figs/smartann.png}{%
        \includegraphics[width=\textwidth]{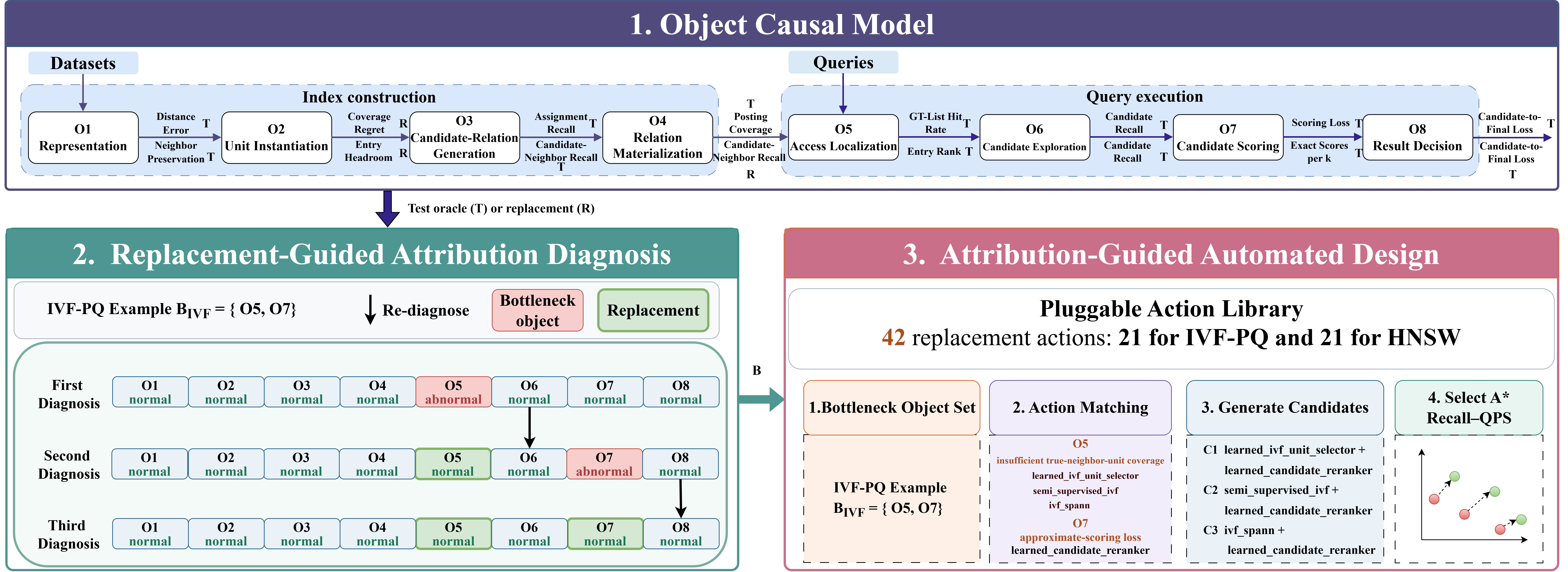}%
    }{%
        \fbox{\parbox[c][1.25in][c]{\textwidth}{\centering
        }}%
    }
    \caption{SmartANN Overview: ANN object causal modeling, bottleneck diagnosis, and automated design.}
    \label{fig:smartann}
\end{figure*}

\section{SmartANN Overview}

As shown in Figure~\ref{fig:smartann}, SmartANN is an object-level framework for replacement-guided diagnosis and auto-design. Its end-to-end workflow consists of an object causal model, an attribution diagnoser, and an auto-designer.

\textbf{\textit{(1) Object causal model.}}
SmartANN instantiates an OCM for the evaluation system defined by the workflow of ANN index construction and query execution. It represents this system as eight ordered objects \(O_1,\ldots, O_8\), with causal effects propagating from upstream objects to downstream objects through their execution behavior and measurable outputs. For a given dataset and query workload, each object defines an input-output contract, expected behavior, and expected output. The contract provides a stable interface for object-level replacement, while the expected behavior and output provide references for evaluating the resulting behavior, output, and downstream effects. A deterministic expected output serves directly as a test oracle; when no exact output can be computed, SmartANN uses a stronger replacement implementation under the same input and resource constraints. This distinction is necessary for heuristic phases such as data partitioning and graph construction.

\textbf{\textit{(2) Attribution diagnoser.}}
SmartANN diagnoses the objects in execution order. The first object \(O_i\) encountered along the execution path that deviates from its expected behavior and output is recorded as a bottleneck, and its output is replaced with the test-oracle output, or its implementation is replaced with a stronger replacement implementation, to suppress loss propagation from \(O_i\) to downstream objects. Once the expected behavior and output at \(O_i\) are achieved, diagnosis resumes from \(O_{i+1}\). This process repeats through \(O_8\) and returns the complete bottleneck set and failure causes.

\textbf{\textit{(3) Auto-designer.}}
The auto-designer comprises a pluggable action library, candidate generation, and design selection. The library contains 42 actions, with 21 for each ANN family. During knowledge preparation, each action is registered with its target objects, failure-cause labels, supported ANN settings, input--output contracts, resource requirements, and compatibility constraints, and is implemented through the common interface. Given a diagnosed bottleneck-object set and its failure causes, SmartANN first retains compatible action compositions that jointly cover the set and then filters them by cause-label matching. It evaluates the resulting designs on the selection split and selects either the highest-Recall design or the highest-QPS design at comparable Recall. Section~\ref{autodesign} provides further details.

\section{Object Causal Model}~\label{decomposition}

The object causal model (OCM) provides SmartANN with the methodological foundation for tracing performance-loss propagation and isolating root causes through controlled object replacement. To support reliable diagnosis and automated design, SmartANN treats an execution phase as a separate object only when it is measurable for diagnosis, controllable through its output during replacement, and replaceable through a fixed contract for automated design. 
Based on the above principles, SmartANN decomposes ANN index construction and query execution into eight ordered objects. Each object is represented as
\begin{equation}
  O_i=\langle M_i,P_i,A_i\rangle .
  \label{eq:object-definition}
\end{equation}
$M_i$ defines its basic function. $P_i$ contains its diagnostic metrics to measure its output quality. $A_i$ contains auto-design actions. An implementation that changes several objects is represented as an \textit{action bundle}. 
Section~\ref{object decomposition} follows the ANN construction and query workflow to define the eight ordered objects and their mechanisms. Section~\label{instantiation} instantiates these objects for IVF-PQ and HNSW and presents their corresponding diagnostic metrics and replacement methods.
\subsection{Object-level ANN Decomposition}~\label{object decomposition}

Based on the OCM, SmartANN decomposes the ANN workflow into the following eight ordered objects and their mechanisms as shown in Figure~\ref{fig:smartann}.

\noindent\underline{\textbf{(1) Representation.}}
This object transforms base and query vectors into the representations used by the ANN algorithm and provides the corresponding distance-computation interface.

\noindent\underline{\textbf{(2) Unit Instantiation.}}
This object creates the basic search units and assigns their structural attributes before candidate relations are generated.

\noindent\underline{\textbf{(3) Candidate-Relation Generation.}}
This object generates candidate relations between base vectors and search units or among search units for subsequent selection and storage.

\noindent\underline{\textbf{(4) Relation Materialization.}}
This object selects and stores candidate relations under structural and resource constraints. When required, it also encodes the associated vector data to produce the searchable structure.

\noindent\underline{\textbf{(5) Access Localization.}}
This object ranks search units or selects entry points for a query, determining where candidate exploration begins.

\noindent\underline{\textbf{(6) Candidate Exploration.}}
This object explores the selected search units and materialized relations within a search budget and produces a candidate vector set.

\noindent\underline{\textbf{(7) Candidate Scoring.}}
This object computes comparable exact or approximate scores for the explored candidates and produces a scored candidate set.

\noindent\underline{\textbf{(8) Result Decision.}}
This object applies deduplication, optional reranking, and top-$k$ selection to produce the final ordered results.

Our division makes each object measurable, controllable, and replaceable. Existing component-level analyses instead divide graph algorithms by design choices~\cite{wang-survey-vldb2021} or IVF-PQ query execution by hardware kernels~\cite{fanns}. Such division supports component comparison, but does not provide the diagnostic metrics, controllable outputs, and replacement implementation required by SmartANN. They therefore cannot be directly reused. We use eight objects because a coarser decomposition would mix different causes of performance loss, whereas a finer decomposition would no longer provide shared interfaces for both IVF-PQ and HNSW.

\subsection{Object Instantiation, Diagnostic Metrics, and Replacement}
\label{instantiation}

SmartANN instantiates IVF-PQ and HNSW as eight ordered objects. For each object, SmartANN specifies its role in each index family, the diagnostic metrics used to detect deviations from expected behavior or output, and a test oracle or stronger replacement implementation for replacement-guided attribution. A test oracle is used when the object has a deterministic expected output. A stronger replacement implementation is used when no unique correct structure exists, with the input, object contract, and resource budget held fixed.

Let $Q_d$ denote the diagnostic query set and let $N_k(q)$ denote the ground-truth top-$k$ neighbor set for query $q$. All object-level quality metrics are measured under a fixed upstream state and comparison budget.

\noindent\textbf{\textit{(1) Representation.}}
This object forms the vector representation and distance-computation interface used by the ANN index. IVF-PQ uses residual representations and a PQ code space. HNSW uses raw vectors or an optional quantized representation. SmartANN measures distance error, rank agreement, and representation-transformation time to characterize representation distortion and its cost. Distance error compares approximate distances with exact distances over the original floating-point vectors. Rank agreement measures the consistency between the candidate orders induced by the two distance functions. Exact distances over the original floating-point vectors form the test oracle. The resulting reference is determined by the distance function, vector-normalization rule, and numerical precision.

\noindent\textbf{\textit{(2) Unit Instantiation.}}
This object instantiates the basic search units. IVF-PQ generates coarse centroids. HNSW assigns level attributes to nodes. For IVF-PQ, clustering error and coverage radius characterize how well the units represent the data space. The inverted-list load factor further measures load imbalance under fixed assignment and materialization rules. Let $L_j$ denote the $j$th inverted list and let $N$ denote the number of base vectors. The load factor is
\begin{equation}
LF=\frac{\max_j |L_j|}{N/\texttt{nlist}}.
\end{equation}
A value of $LF$ closer to one indicates more balanced search units. A larger value indicates that a small number of units contain disproportionately many vectors, which may increase query-time scanning cost. For HNSW, the level distribution and top-layer size characterize the hierarchy. SmartANN further measures the mean number of upper-layer navigation hops. Let $h_q$ denote the number of upper-layer hops for query $q$. This metric is
\begin{equation}
H_2=\frac{1}{|Q_d|}\sum_{q\in Q_d}h_q.
\end{equation}
A smaller $H_2$ indicates more direct hierarchical navigation. Neither search-unit formation nor hierarchy construction has a unique correct output. SmartANN therefore selects a stronger replacement implementation under a fixed unit cardinality, construction budget, memory budget, and randomization protocol.

\noindent\textbf{\textit{(3) Candidate-Relation Generation.}}
This object generates candidate relations for subsequent materialization. IVF-PQ generates candidate assignments between vectors and coarse centroids. HNSW generates candidate neighbors during node insertion. SmartANN uses candidate-relation Recall to measure the completeness of the generated relations relative to the conditional exact relations. Distance gap and distance-computation count characterize relation quality and generation cost. Let $E_3$ denote the generated relation set and let $E_3^*$ denote the reference set obtained through exhaustive distance computation under the fixed upstream state. Candidate-relation Recall is
\begin{equation}
R_3=\frac{|E_3\cap E_3^*|}{|E_3^*|}.
\end{equation}
Once the centroids or the HNSW insertion prefix are fixed, exhaustive distance computation yields a conditional test oracle. Its output also depends on the distance function and candidate-relation cardinality.

\noindent\textbf{\textit{(4) Relation Materialization.}}
This object selects and stores candidate relations under structural and resource constraints. IVF-PQ writes postings and PQ codes according to the existing assignments. HNSW materializes the selected relations as hierarchical adjacency edges. For IVF-PQ, posting coverage and quantization error measure storage completeness and encoding distortion. For HNSW, node-degree distribution, graph connectivity, reachability, and path detour characterize structural constraints, global connectivity, and graph navigability. Multiple relation-selection policies can produce valid structures, so this object has no unique test oracle. SmartANN fixes the candidate relations, degree constraints, index size, and construction budget, then selects a stronger replacement implementation with better structural quality.

\noindent\textbf{\textit{(5) Access Localization.}}
This object determines where a query starts accessing the index. IVF-PQ ranks coarse units and selects the $\texttt{nprobe}$ inverted lists. HNSW selects a base-layer entry point. For IVF-PQ, true-neighbor-unit coverage measures how completely the selected units cover the ground-truth neighbors. Let $S_q$ denote the unit set selected for query $q$ and let $u(x)$ denote the inverted-list unit containing vector $x$. The per-query coverage and its mean are
\begin{equation}
R_5(q)=\frac{1}{k}\sum_{x\in N_k(q)}\mathbb{I}[u(x)\in S_q],
\qquad
\bar{R}_5=\frac{1}{|Q_d|}\sum_{q\in Q_d}R_5(q).
\end{equation}
A larger $\bar{R}_5$ indicates that Access Localization preserves more complete sources of ground-truth neighbors for downstream exploration. For HNSW, SmartANN measures entry-point hit rate, entry-distance rank, and ground-truth reachability under a fixed downstream search budget. Given a fixed output cardinality, exact unit ranking for IVF-PQ or the best feasible entry for HNSW forms a conditional test oracle. The concrete reference is jointly determined by the distance function, $\texttt{nprobe}$ or entry cardinality, upstream index structure, and downstream search budget.

\noindent\textbf{\textit{(6) Candidate Exploration.}}
This object enumerates the selected postings or traverses the HNSW base layer within a search budget. It outputs a candidate vector set $C_q$. Its primary quality metric is candidate Recall
\begin{equation}
R_6(q)=\frac{|C_q\cap N_k(q)|}{k},
\qquad
\bar{R}_6=\frac{1}{|Q_d|}\sum_{q\in Q_d}R_6(q).
\end{equation}
SmartANN also measures the 10th percentile of candidate Recall, denoted by $R_6^{\mathrm{p10}}$, and the fraction of queries whose candidate Recall falls below a target $\tau_6$
\begin{equation}
\rho_6=\frac{1}{|Q_d|}\sum_{q\in Q_d}\mathbb{I}[R_6(q)<\tau_6].
\end{equation}
The mean, 10th percentile, and below-target fraction capture overall candidate coverage, tail quality on difficult queries, and the scope of the abnormality. The numbers of scanned postings, visited nodes, candidate expansions, and distance computations, together with execution time, measure exploration cost. Under a fixed upstream output, exhaustive enumeration or reachable-set traversal provides a coverage oracle. Cost-sensitive diagnosis instead uses a budget-matched reference. The reference depends on the selected units or entry point, graph reachability, search budget, and stopping condition.

\noindent\textbf{\textit{(7) Candidate Scoring.}}
This object computes comparable exact or approximate scores over a fixed candidate set. Let $R_7(q)$ denote the fraction of ground-truth top-$k$ neighbors retained at the scoring boundary. SmartANN defines scoring loss as
\begin{equation}
L_7(q)=R_6(q)-R_7(q),
\qquad
\bar{L}_7=\frac{1}{|Q_d|}\sum_{q\in Q_d}L_7(q).
\end{equation}
A larger $\bar{L}_7$ indicates more severe loss during approximate distance computation or score-based ordering. Distance error, rank agreement, and order-flip rate further explain the source of this loss. The number of scored candidates, distance-computation count, and scoring time characterize its cost. Exact distances over the same candidate identifiers form the test oracle. This reference is determined by the distance function, vector representation, and numerical precision. Exact reranking over a fixed candidate set also belongs to this object.

\noindent\textbf{\textit{(8) Result Decision.}}
This object applies deduplication, tie handling, and top-$k$ selection to the existing candidates and scores. It produces the final ordered result. Let $R_8(q)$ denote the ground-truth-neighbor coverage of the final result. Candidate-to-final loss is
\begin{equation}
L_8(q)=R_7(q)-R_8(q).
\end{equation}
This metric measures the additional loss introduced when the output of Candidate Scoring is converted into the final result. Selection agreement, top-$k$ agreement, and decision time characterize result consistency and decision cost. When candidates, scores, the deduplication policy, and the tie-breaking rule are fixed, deterministic top-$k$ selection forms the test oracle. Its output is determined by $k$ and the tie-handling policy.

\subsection{Running example}
We use GloVe as a running example throughout Sections~\ref{diagnosis} and~\ref{autodesign} to illustrate the complete SmartANN workflow for IVF-PQ and HNSW. Given the same datasets and queries, the attribution diagnoser identifies different bottleneck-object sets and failure causes for the two ANN families. For IVF-PQ, it identifies Access Localization \(O_5\) and Candidate Scoring \(O_7\), caused by insufficient true-neighbor-unit coverage and approximate-scoring loss, respectively. For HNSW, it identifies Candidate Exploration \(O_6\), caused by insufficient candidate coverage. The auto-designer first filters candidate plans by their coverage of the diagnosed bottleneck-object set and then matches their actions against the diagnosed failure causes. This process generates a unit-selection and reranking design for IVF-PQ and a safe search-expansion design for HNSW.

\section{Replacement-Guided Attribution Diagnosis}~\label{diagnosis}

\begin{algorithm}[t]
\caption{Sequential object-level bottleneck diagnosis}
\label{alg:object-diagnosis}
\begin{algorithmic}[1]

\Require Ordered objects \(O_1,\ldots,O_8\)
\Ensure Bottleneck object set \(\mathcal{B}\)

\State \(\mathcal{B} \gets \varnothing\)

\For{\(i \gets 1\) \textbf{to} \(8\)}

    \State Diagnose \(O_i\) using \(P_i\)

    \If{\(P_i\) is abnormal}

        \State \(\mathcal{B}
            \gets
            \mathcal{B}\cup\{O_i\}\)

        \If{a test oracle exists for \(O_i\)}
            \State \(O_i^\star
                \gets
                \operatorname{TestOracle}(O_i,A_i)\)
        \Else
            \State \(O_i^\star
                \gets
                \operatorname{Replacement}(O_i,A_i)\)
        \EndIf

        \State Replace \(O_i\) with \(O_i^\star\)
        \State Re-execute \(O_i,\ldots,O_8\)
        \State \textbf{assert} \(P_i\) is normal
        \State Retain \(O_i^\star\)

    \EndIf

\EndFor

\State \Return \(\mathcal{B}\)

\end{algorithmic}
\end{algorithm}

As shown in Algorithm~\ref{alg:object-diagnosis}, SmartANN identifies the bottleneck-object set through a sequential \textit{diagnose-and-replace} loop, in which it diagnoses the first deviating object, replaces its output using the test oracle or a stronger replacement implementation, and diagnosis continues from the subsequent object until \(O_8\). 

\textbf{\textit{Step 1: Object-level diagnosis.}}
SmartANN examines \(O_1\) through \(O_8\) in execution order and stops at the first object \(O_i\) whose diagnostic metrics in \(P_i\) deviate from their expected outputs. This object is recorded as a bottleneck. Because the output of \(O_i\) is consumed by subsequent objects, its performance loss may either propagate downstream or alter downstream inputs and mask independent bottlenecks. SmartANN therefore controls \(O_i\) before diagnosing subsequent objects.

\textit{We return to the \underline{\textbf{running example}} to illustrate the first round of object-level diagnosis.}
or IVF-PQ, \(O_1\)--\(O_4\) remain within their expected ranges,
whereas Access Localization \(O_5\) has an average
true-neighbor-unit coverage of only \(\bar{R}_5=0.644\).
SmartANN therefore records \(O_5\) as the first bottleneck and
suspends diagnosis of \(O_6\)--\(O_8\). For HNSW, \(O_1\)--\(O_5\)
remain within their expected ranges, whereas Candidate Exploration
\(O_6\) has a mean candidate Recall of \(\bar{R}_6=0.863\) and a
10th-percentile Recall of \(R_6^{\mathrm{p10}}=0.570\).
SmartANN records \(O_6\) as the first bottleneck and suspends
diagnosis of \(O_7\) and \(O_8\). These results determine the
objects replaced in Step~2.

\textbf{\textit{Step 2: Bottleneck replacement.}}
SmartANN selects a replacement implement for \(O_i\) from \(A_i\). If \(O_i\) has a deterministic expected output, SmartANN uses it as a test oracle. Examples include exact distance computation over fixed candidate identifiers and exact top-\(k\) selection over fixed scores. If \(O_i\) has a deterministic expected output, SmartANN replaces its output with the test oracle. Otherwise, SmartANN replaces its implementation with a stronger replacement implementation under the same inputs and resource constraints. Among the available implementations that satisfy the expected behavior of \(O_i\), SmartANN selects the one with the least measured adverse effect on downstream objects, and treats its output as the expected output of \(O_i\).

Let \(O_i^\star\) denote the replaced object. SmartANN replaces \(O_i\) with \(O_i^\star\) and re-executes \(O_i\) and the affected downstream object while keeping the actions of other objects unchanged. This controlled execution attributes the observed downstream changes \(p_j\in P_j\) to the replacement of \(O_i\).
The replacement operator is defined as:

\begin{equation}
\operatorname{rep}
\left(O_j; O_i \rightarrow O_i^\star\right)
=
\left|p_j(O_i^\star)-p_j(O_i)\right|.
\label{eq-replacement-operator}
\end{equation}

A non-zero value indicates that the replacement affects \(O_j\). Whe-\\ther the effect is beneficial or adverse is determined by the semantics and values of \(p_j\). Changes within the measurement tolerance are treated as zero.

\textit{We return to the \underline{\textbf{running example}} to quantify replacement operator effects.}
For IVF-PQ, replacing \(O_5\) increases its true-neighbor-unit coverage from 0.644 to 1.000, allowing more true-neighbor candidates to reach \(O_7\). The average scoring loss at \(O_7\) consequently increases from 0.184 to 0.421. The replacement operator gives
$
\operatorname{rep}_{\bar{L}_7}
\left(O_7;O_5\rightarrow O_5^\star\right)
=
|0.421-0.184|
=
0.237.
$
Because a lower scoring loss is preferred, this change reveals a Candidate Scoring bottleneck previously masked by incomplete access localization. Replacing \(O_7\) then reduces its scoring loss to zero.

For HNSW, replacing \(O_6\) increases average Candidate Recall from 0.863 to 0.946 and its 10th-percentile value from 0.570 to 0.860, while reducing the fraction of below-target queries from 35\% to 1\%. Candidate Scoring introduces no additional loss under the updated candidate input. The downstream deviation is therefore attributed to performance-loss propagation from \(O_6\), rather than an independent \(O_7\) bottleneck.

\textbf{\textit{Step 3: Repeated diagnosis and replacement.}}

Once \(P_i\) returns to its expected range, SmartANN retains \(O_i^\star\) and resumes diagnosis from \(O_{i+1}\). Previously retained replacements prevent confirmed upstream losses from affecting subsequent diagnosis. This \textit{diagnose-and-replace} loop repeats until \(O_8\) has been examined. All objects identified through their own abnormal diagnostic metrics form the complete bottleneck-object set and failure causes.

\textit{We return to the \underline{\textbf{running example}} to illustrate how SmartANN identifies the complete bottleneck-object set and the corresponding failure causes through sequential diagnosis and replacement.}

For IVF-PQ, the diagnostic metrics of \(O_1\)--\(O_4\) remain
within their expected ranges. Access Localization \(O_5\) is the
first deviating object, with an average true-neighbor-unit coverage
of only \(\bar{R}_5=0.644\), indicating insufficient access to the
units containing true neighbors. SmartANN replaces its output with
the test-oracle output and re-executes \(O_6\)--\(O_8\). Under the
updated execution, \(O_6\) satisfies its diagnostic conditions,
whereas the average scoring loss at Candidate Scoring \(O_7\)
increases to \(\bar{L}_7=0.421\), exposing approximate-scoring loss
as an independent failure cause. SmartANN therefore records
\(O_7\) as a bottleneck and replaces it with exact distance
computation over the same candidate identifiers. The diagnosis
returns
$
\mathcal{B}_{\mathrm{IVF}}=\{O_5,O_7\},
$
where \(O_5\) exhibits insufficient true-neighbor-unit coverage and
\(O_7\) exhibits approximate-scoring loss.

For HNSW, \(O_1\)--\(O_5\) remain within their expected ranges.
Candidate Exploration \(O_6\) is the first deviating object, with
an average candidate Recall of \(\bar{R}_6=0.863\) and a
10th-percentile Recall of \(R_6^{\mathrm{p10}}=0.570\). These
metrics indicate insufficient candidate exploration, particularly
for difficult queries. After SmartANN replaces its candidate output
with the test-oracle output, Candidate Scoring \(O_7\) and Result
Decision \(O_8\) satisfy their diagnostic conditions. The diagnosis
therefore returns
$
\mathcal{B}_{\mathrm{HNSW}}=\{O_6\},
$
with insufficient candidate coverage as the failure cause of
\(O_6\).

\section{Attribution-Guided Auto-Design}
\label{autodesign}

Given the diagnosed bottleneck-object set, the auto-designer first automatically constructs candidate designs from individual actions or compatible action compositions whose joint coverage spans the complete set. SmartANN then instantiates and evaluates each candidate and selects the best-performing design according to the specified performance objective. This automated process comprises candidate generation, candidate evaluation, and selection.

\subsection{Pluggable Action Library}




\noindent\textbf{(1) Knowledge preparation.}
SmartANN registers an ANN optimization as an action only when the optimization can be mapped to one or more objects, and its applicability conditions can be specified explicitly. Each action record declares its supported ANN family, covered objects, failure causes, action type, training requirement, index-reconstruction requirement, budget requirement, compatibility mode, and output contract. The covered objects determine which bottlenecks the action can repair. The failure cause further distinguishes optimization directions for the same object. For example, insufficient candidate Recall at Candidate Exploration $O_6$ calls for revised exploration. When candidate Recall already meets the target, but access cost remains excessive, adaptive termination or distance-computation optimization is more appropriate.

SmartANN registers an action record for each action in the library. We illustrate the record schema using \texttt{hnsw\_adaptive\_entry}, an HNSW action bundle that jointly replaces two objects. This example shows how an action record specifies the target objects and the corresponding performance bottlenecks that the action is designed to address.

\begin{center}
\setlength{\fboxsep}{6pt}
\fcolorbox{black}{white}{%
\begin{minipage}{0.92\linewidth}
\footnotesize
\renewcommand{\arraystretch}{1.18}

\begin{tabularx}{\linewidth}{
@{}
>{\bfseries\raggedright\arraybackslash}p{0.38\linewidth}
>{\raggedright\arraybackslash}X
@{}
}
\rowcolor{LightCyan}
\multicolumn{2}{c}{\textbf{Action Record}} \\
\addlinespace[3pt]

Action
& \texttt{hnsw\_adaptive\_entry} \\

Applicable ANN family
& HNSW \\

Covered objects
& $\{O_5,O_6\}$ \\

Label
& Repairs poor entry quality and downstream candidate-coverage loss \\

Type
& Data-driven \\

Training and reconstruction
& Neither required \\

Budget
& Fixed baseline \texttt{efSearch} with all additional costs included \\

Compatibility
& Direct query-side replacement \\

Output contract
& Entry and candidate-set boundary \\

\end{tabularx}
\end{minipage}%
}
\end{center}

\noindent\textbf{(2) Action Implementation.}
SmartANN defines a standardized input--output interface for each object and requires all actions targeting that object to conform to it. Each action is implemented as a pluggable action, allowing heterogeneous optimizations to be interchanged and composed without modifying adjacent objects. The library contains 42 actions, of which 34 are adapted from 28 existing ANN optimization methods, while eight are implemented specifically for SmartANN. These actions are evenly divided between IVF-PQ and HNSW, with 21 for each family. By implementation type, 17 are learned, 12 are data-driven, and 13 are non-learned.

SmartANN implements its action runtime in C++17 and defines a standardized input--output interface for each object. All actions targeting the same object conform to this interface and are implemented as pluggable actions. Each action provides a common \texttt{prepare} interface for constructing or loading action-specific models and index artifacts, and an \texttt{execute} interface for consuming object inputs and producing contract-compliant outputs. Learned actions are trained offline on the \emph{learned training} split, and their resulting models are loaded by the corresponding C++ actions. During execution, SmartANN validates action outputs, resource usage, and evidence observed at downstream object boundaries.



\noindent\textbf{(3) Diagnostic replacement and auto-design actions.}
Diagnostic replacement and auto-design replacement serve different purposes. Diagnostic replacement uses a test oracle or a stronger replacement implementation to suppress performance-loss propagation to downstream objects. Such replacement may incur prohibitive computational cost or require information unavailable during online query processing. They are therefore unsuitable as deployable redesigns. Auto-design replacement instead uses deployable design actions selected from attribution evidence. These actions improve the Recall--QPS tradeoff under the object contracts and resource constraints. SmartANN prefers higher QPS at comparable Recall and higher Recall when the remaining conditions are comparable.

\subsection{Design Generation and Selection}
Let \(B\) denote the diagnosed bottleneck-object set. SmartANN uses the diagnostic metrics to determine the bottleneck cause of each object in \(B\), and then applies two-stage filtering over the action library. The first stage retains only individual actions or compatible action compositions that jointly cover every object in \(B\), eliminating candidates unrelated to the diagnosed bottlenecks. The second stage matches the failure cause of each bottleneck with the failure causes addressed by the retained actions. A candidate is retained if it contains at least one action that matches the diagnosed failure cause of a corresponding bottleneck object in \(B\). This process produces a small set of relevant candidate designs.

SmartANN materializes and evaluates the filtered candidates using a small, dedicated selection query set. A candidate containing actions for \(O_1\)--\(O_4\) requires the affected index structures to be rebuilt; the corresponding library actions support parallel cluster and graph construction to reduce this overhead. A candidate containing only actions for \(O_5\)--\(O_8\) reuses the existing index and parallelizes the selection queries. Depending on the optimization objective, SmartANN selects either the design with the highest Recall or the design with the highest QPS at matched Recall. The selected design is finally evaluated on the final evaluation queries detailed in Section.

\textit{We return to the \underline{\textbf{running example}} to illustrate the automated redesign process.}
The auto-designer consumes the bottleneck-object sets and failure causes produced by attribution diagnosis. Each action record specifies its target objects and a failure-cause label describing the bottleneck that the action addresses. The first stage filters
candidate designs by object coverage, whereas the second stage matches the diagnosed failure causes against these registered labels.
A candidate that passes the coverage filter is retained if at least one of its actions has a label matching a diagnosed failure cause.

For IVF-PQ, the first stage retains only individual actions or compatible action compositions that jointly cover
$
\mathcal{B}_{\mathrm{IVF}}=\{O_5,O_7\}.
$
The diagnosed cause at \(O_5\), insufficient true-neighbor-unit coverage, matches the corresponding label of unit-selection actions.
Similarly, the approximate-scoring loss at \(O_7\) matches the label of candidate-reranking actions. This matching retains a small set of plans containing at least one cause-matched action. After evaluation
on the selection split, the selected composition combines unit selection and candidate reranking, improving Recall@100 from 0.2670
to 0.6543.

For HNSW, the first stage retains the actions targeting
$
\mathcal{B}_{\mathrm{HNSW}}=\{O_6\},
$
including \texttt{hnsw\_safe\_search\_expansion},
\texttt{hnsw\_laet}, \texttt{hnsw\_\\adsampling}, and
\texttt{hnsw\_dade}. The diagnosed cause, insufficient candidate
coverage, matches the failure-cause label registered for \texttt{hnsw\_safe\_search\_expansion}. The remaining actions are labeled for reducing exploration or distance-computation cost and are therefore pruned. The selected design improves Recall@100 from
0.6692 to 0.8162.

This paired example shows that the same dataset and query workload lead to ANN-family-specific bottleneck-object sets, failure causes, and automated redesigns.

\section{Evaluation}
\label{sec:evaluation}
This section conducts an extensive evaluation of SmartANN against the state-of-the-art methods on eight real-world datasets using IVF-PQ and HNSW.
All experiments run on a server equipped with two 2.0 GHz Intel Xeon Gold 6330 processors, 56 physical cores, 112 logical cores, and 503 GiB of memory. The server runs Ubuntu 20.04. All implementations are compiled with GCC 9.4.0. All reported query experiments use 28 logical cores with hyper-threading enabled

\subsection{Experimental Setup}
\label{sec:eval-setup}

\subsubsection{Datasets and Query Splits}

\noindent\textbf{ (1) Datasets.}
We evaluate SmartANN on eight real-world datasets. SIFT~\cite{5432202}, Deep~\cite{7780595}, GIST~\cite{5432202,oliva2001modeling}, and Tiny5M~\cite{torralba200880} represent image retrieval, while GloVe~\cite{pennington2014glove}, MSong~\cite{bertin2011million}, Yandex-200~\cite{simhadri2022results}, and Wikipedia~\cite{bvb,girdhar2023imagebind} contain word, audio, search, and cross-modal embeddings, respectively. Hubness-GloVe, Small-Gap-Deep, Density-LID-GIST, and OOD-Wikipedia further emphasize hubness, small nearest-neighbor gaps, heterogeneous local density, and distribution shift. The datasets contain 0.48--9.99 million base vectors with 96--1,024 dimensions and cover L2, angular, and cosine distances. Because the Wikipedia slice contains many duplicate vectors, we analyze it separately.

\textbf{ (2) Query splits.}
Using a fixed random seed, we partition the queries into five mutually disjoint splits. The \emph{tuning} split constructs the baseline Recall--QPS Pareto frontier, the \emph{diagnostic} split supports object diagnosis and replacement, the \emph{learned training} split trains learned actions, the \emph{selection} split selects candidate designs and configurations, and the \emph{final evaluation} split reports final performance. The final evaluation queries are never used for parameter tuning, diagnosis, learned training, or candidate selection, preventing test-set contamination and ensuring an unbiased evaluation.

To demonstrate that SmartANN improves both query effectiveness and efficiency beyond a single point, we select three representative points from the baseline Pareto frontier. They include the maximum-Recall point, the point maximizing $\mathrm{Recall}\times\mathrm{QPS}$, and the point with the largest minimum Recall distance from the first two. These well-separated representative points serve as the starting configurations for diagnosis and automated redesign.

\subsubsection{Evaluation Methodology}
To the best of our knowledge, no existing work jointly supports bottleneck attribution and automated ANN redesign. We therefore compare SmartANN with VDTuner~\cite{yang2024vdtuner}, a state-of-the-art ANN parameter-tuning framework. To isolate the effectiveness of bottleneck attribution, we also implement a reproducible Random baseline using fixed random seeds.

\noindent\textbf{(1) Parameter spaces.}
We construct the baseline parameter space using parameters that directly determine the index structure and query cost. Across the eight datasets, the evaluated IVF-PQ values are $nlist\in\{128,256,512,1024\}$, $m\in\{10,14,16,20,32\}$, $nbits\in\{6,8\}$, $nprobe\in\{8,16,32,64\}$, and exact-reranking factors from $\{2,4,8,16\}$. We retain only dimension-compatible values of $m$.

For HNSW, the evaluated values are $M\in\{12,16,24\}$, $efCons-\\truction\in\{160,200,240,300\}$, and $efSearch\in\{100,128,160,192,\\256,384,576\}$. Candidate-exploration and scoring costs are governed by $efSearch$. For both ANN families, we use dataset-specific subsets of these values to cover efficiency-oriented and quality-oriented Recall--QPS regions. SmartANN generates candidate redesigns from 42 registered actions, with 21 actions for IVF-PQ and 21 actions for HNSW.

\noindent\textbf{(2) SmartANN.}
For each frozen representative point, SmartANN runs the \textit{diagnose-and-replace} loop on the \texttt{diagnostic} split to obtain the complete bottleneck-object set and its failure causes. It filters the action library by bottleneck coverage, failure-cause matching, interface compatibility, and resource constraints. Learned actions use only the \texttt{learned training} split, and the resulting candidate designs are evaluated on the \texttt{selection} split. The selected design and its parameters are then frozen for \texttt{final evaluation}.

\noindent\textbf{(3) Random baseline.}
Random follows the same action-training, selection, and final-evaluation protocol but cannot access diagnostic metrics or the bottleneck-object set and failure causes. For HNSW, it samples \(K=3\) distinct action--budget combinations without replacement using fixed seeds 14, 28, and 42 and budget multipliers from \(\{0.5,0.75,1.0\}\).

\noindent\textbf{(4) VDTuner.}
VDTuner performs multi-objective black-box search over the complete parameter space on the \texttt{tuning} split. 
For a fair comparison with SmartANN's three redesigns, we select the three VDTuner configurations with the highest Recall on the \texttt{tuning} split and freeze them before final evaluation.

\subsubsection{Ablation Methodology}

We ablate two key designs of SmartANN, namely attribution-guided auto-design generation using the bottleneck-object set and failure causes, as well as sequential replacement for downstream attribution.

\noindent\textbf{(1) Bottleneck-object set.}
We compare SmartANN with Random under the same baseline representative point, action library, training and selection data, resource budget, and final evaluation protocol. SmartANN selects designs using the diagnosed bottleneck-object set and failure causes, whereas Random cannot access this attribution evidence and samples three distinct valid designs using fixed seeds 14, 28, and 42. This comparison isolates the benefit of attribution-guided auto-design generation.

\noindent\textbf{(2) Replacement Operator.}
To validate the effect of the replacement operator, we compare Replacement, which replaces \(O_i\) and re-executes the affected downstream objects, with No-Replacement, which retains the original downstream execution, while keeping the object order, diagnostic metrics, and detection thresholds fixed. This comparison evaluates whether the operator can distinguish propagated performance loss from independent or masked downstream bottlenecks. 

\subsection{Performance Comparison}
\label{sec:eval-results}

We first report object-level diagnosis and replacement results for IVF-PQ and HNSW. We then compare the end-to-end Recall--QPS performance of SmartANN, the original baseline, and VDTuner. We finally analyze the costs of diagnosis, training, index construction, and complete-configuration evaluation.

\begin{figure*}[t]
    \centering
    \includegraphics[width=\textwidth]{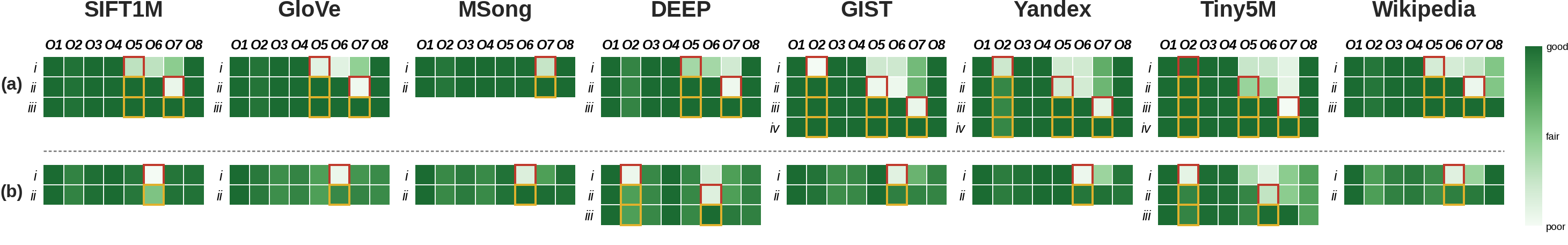}
    \caption{Sequential object diagnosis and replacement at representative operating points for IVF-PQ in panel (a) and HNSW in panel (b). Darker green indicates a better normalized primary diagnostic metric. Red borders mark newly diagnosed bottleneck objects. Yellow borders mark replacements retained from preceding iterations.}
    \label{fig:diagnosis-heatmap}
\end{figure*}

\subsubsection{Diagnosis and Replacement Results}
Figure~\ref{fig:diagnosis-heatmap} shows the sequential attribution process for representative IVF-PQ and HNSW operating points across all datasets. Each row represents one iteration of the \textit{diagnose-and-replace} loop. Cell color indicates how closely each object's normalized primary diagnostic metric approaches its expected range. Red borders mark the first bottleneck identified in the current iteration, whereas yellow borders mark previously replaced objects. Diagnosis terminates when no new bottleneck is found.

Figure~\ref{fig:diagnosis-heatmap}~(a) shows that IVF-PQ bottlenecks mainly occur at Unit Instantiation \(O_2\), Access Localization \(O_5\), and Candidate Scoring \(O_7\), with different workloads producing bottleneck-object sets of different sizes. On GIST, replacing \(O_2\) reduces the inverted-list load factor from 5.547 to 1.813. Replacing \(O_5\) then increases true-neighbor-unit coverage from 0.885 to 1.000, while the scoring loss at \(O_7\) rises from 0.242 to 0.338, exposing a previously masked scoring bottleneck. Replacing \(O_7\) reduces this loss to zero, yielding the bottleneck-object set \(\{O_2,O_5,O_7\}\).

Figure~\ref{fig:diagnosis-heatmap}~(b) shows that HNSW bottlenecks mainly occur at Unit Instantiation \(O_2\) and Candidate Exploration \(O_6\). On GloVe, replacing \(O_6\) improves the mean and 10th-percentile candidate Recall from 0.863 and 0.570 to 0.946 and 0.860, respectively, while reducing the fraction of queries below the target from 35\% to 1\%. These results attribute the performance loss to insufficient candidate exploration across both typical and difficult queries.

Tiny5M contains bottlenecks at both \(O_2\) and \(O_6\). Replacing \(O_2\) reduces the mean upper-layer navigation hops from 10.28 to 8.31, but improves mean candidate Recall only from 0.714 to 0.735. Replacing \(O_6\) further increases it to 0.829 and reduces the fraction of below-target queries to 1\%. SmartANN therefore identifies \(O_2\) and \(O_6\) as independent bottlenecks, while \(O_7\) and \(O_8\) introduce no additional quality loss. Here, all candidate Recall values are measured on the diagnostic queries rather than from final evaluation queries.

\subsubsection{End-to-End Performance }

\begin{figure}[t]
    \centering
    \includegraphics[width=\columnwidth]{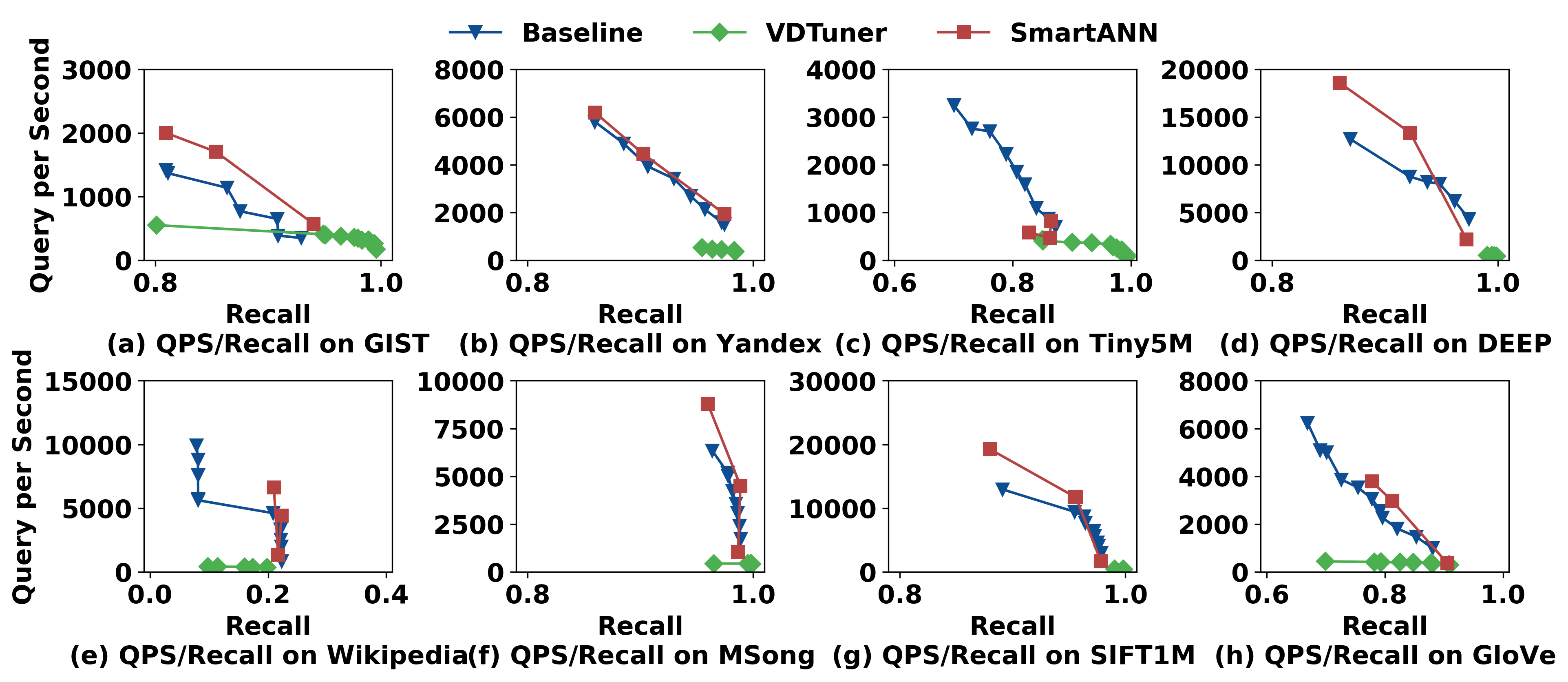}
    \caption{Recall--QPS comparison among the HNSW baseline, VDTuner, and SmartANN.}
    \label{fig:hnsw-end-to-end}
\end{figure}

Figure~\ref{fig:hnsw-end-to-end} and Table~\ref{tab:ivfpq-qps-recall} report the end-to-end results for HNSW and IVF-PQ, respectively. When retrieval quality is the primary limiting factor, SmartANN selects a quality-oriented redesign. At IVF-PQ GloVe-P1, for example, repairing Access Localization $O_5$ and Candidate Scoring $O_7$ increases Recall from 0.2670 to 0.6543.

When candidate Recall already satisfies the requirement but vector accesses or distance computations remain redundant, SmartANN selects an efficiency-oriented action. At the high-Recall HNSW operating point on Yandex-200, SmartANN selects AdSampling. It preserves Recall at 0.9747 and increases QPS from 1,506 to 1,939. SmartANN retains the original design when diagnosis finds no bottleneck.

SmartANN produces more favorable Recall--QPS tradeoffs than VDTuner on most datasets and operating regions. VDTuner jointly tunes construction and query parameters through black-box search. High-performance solutions often require aggressive construction and search configurations. Every change to a construction parameter also requires index reconstruction and reevaluation. VDTuner consequently incurs a large search space and high iteration cost. SmartANN instead restricts redesign to actions that match confirmed bottleneck objects. This restriction reduces ineffective complete-design evaluations.

\begin{table}[!t]
  \footnotesize
  \centering
  \setlength{\tabcolsep}{4pt}
  \caption{IVF-PQ QPS and Recall for the baseline, VDTuner, and SmartANN on eight datasets at three operating points. The best Recall and QPS in each row are shown in bold.}
  \begin{tabular}{@{}cccc|cc|cc@{}}
  \toprule[1.5pt]
  \multirow{2}{*}{Dataset} & \multirow{2}{*}{Point} &
    \multicolumn{2}{c}{Baseline} &
    \multicolumn{2}{c}{VDTuner} &
    \multicolumn{2}{c}{SmartANN} \\
  \cmidrule(lr){3-4}\cmidrule(lr){5-6}\cmidrule(lr){7-8}
   & & \begin{tabular}[c]{@{}c@{}}Recall\end{tabular} & \begin{tabular}[c]{@{}c@{}}QPS\end{tabular} & \begin{tabular}[c]{@{}c@{}}Recall\end{tabular} & \begin{tabular}[c]{@{}c@{}}QPS\end{tabular} & \begin{tabular}[c]{@{}c@{}}Recall \end{tabular} & \begin{tabular}[c]{@{}c@{}}QPS\end{tabular} \\
  \midrule
  \rowcolor{Gray!16} GIST & P1 & $0.3888$ & \textcolor{BestGreen}{$2139.8$} & $0.3958$ & $408.2$ & \textcolor{BestGreen}{$0.3969$} & $873.9$ \\
  \rowcolor{WhiteColr}  & P2 & $0.6599$ & \textcolor{BestGreen}{$1377.7$} & $0.3921$ & $464.5$ & \textcolor{BestGreen}{$0.6638$} & $699.6$ \\
  \rowcolor{LightCyan}  & P3 & $0.8985$ & $256.5$ & $0.3847$ & \textcolor{BestGreen}{$393.8$} & \textcolor{BestGreen}{$0.9038$} & $124.4$ \\
  \midrule
  \rowcolor{Gray!16} Yandex & P1 & $0.2452$ & \textcolor{BestGreen}{$8860.2$} & \textcolor{BestGreen}{$0.3258$} & $472.2$ & $0.2500$ & $1907.4$ \\
  \rowcolor{WhiteColr}  & P2 & $0.4607$ & \textcolor{BestGreen}{$5109.6$} & $0.3254$ & $183.5$ & \textcolor{BestGreen}{$0.4780$} & $1514.5$ \\
  \rowcolor{LightCyan}  & P3 & $0.7250$ & \textcolor{BestGreen}{$1577.9$} & $0.3244$ & $372.1$ & \textcolor{BestGreen}{$0.7405$} & $309.7$ \\
  \midrule
  \rowcolor{Gray!16} Tiny5M & P1 & $0.3753$ & \textcolor{BestGreen}{$649.4$} & $0.3774$ & $352.1$ & \textcolor{BestGreen}{$0.3777$} & $99.14$ \\
  \rowcolor{WhiteColr}  & P2 & $0.6180$ & \textcolor{BestGreen}{$521.6$} & $0.3766$ & $414.2$ & \textcolor{BestGreen}{$0.6258$} & $96.04$ \\
  \rowcolor{LightCyan}  & P3 & $0.8974$ & $73.79$ & $0.3729$ & \textcolor{BestGreen}{$424.2$} & \textcolor{BestGreen}{$0.9123$} & $11.64$ \\
  \midrule
  \rowcolor{Gray!16} DEEP & P1 & $0.7225$ & \textcolor{BestGreen}{$774.8$} & $0.4000$ & $433.0$ & \textcolor{BestGreen}{$0.9436$} & $96.57$ \\
  \rowcolor{WhiteColr}  & P2 & $0.8353$ & \textcolor{BestGreen}{$741.6$} & $0.3999$ & $468.6$ & \textcolor{BestGreen}{$0.9436$} & $95.96$ \\
  \rowcolor{LightCyan}  & P3 & \textcolor{BestGreen}{$0.9637$} & $160.6$ & $0.3992$ & \textcolor{BestGreen}{$323.8$} & \textcolor{BestGreen}{$0.9637$} & $160.6$ \\
  \midrule
  \rowcolor{Gray!16} Wikipedia & P1 & $0.1590$ & \textcolor{BestGreen}{$2091.0$} & $0.1170$ & $391.3$ & \textcolor{BestGreen}{$0.8882$} & $415.1$ \\
  \rowcolor{WhiteColr}  & P2 & $0.1963$ & \textcolor{BestGreen}{$1754.8$} & $0.1168$ & $371.1$ & \textcolor{BestGreen}{$0.9370$} & $378.7$ \\
  \rowcolor{LightCyan}  & P3 & $0.2063$ & \textcolor{BestGreen}{$1036.5$} & $0.1160$ & $422.6$ & \textcolor{BestGreen}{$0.9483$} & $337.2$ \\
  \midrule
  \rowcolor{Gray!16} MSong & P1 & $0.6080$ & \textcolor{BestGreen}{$3470.8$} & $0.5455$ & $390.4$ & \textcolor{BestGreen}{$0.9872$} & $207.2$ \\
  \rowcolor{WhiteColr}  & P2 & $0.7533$ & \textcolor{BestGreen}{$2729.3$} & $0.5454$ & $388.2$ & \textcolor{BestGreen}{$0.9872$} & $207.8$ \\
  \rowcolor{LightCyan}  & P3 & \textcolor{BestGreen}{$0.9736$} & \textcolor{BestGreen}{$1158.6$} & $0.5450$ & $178.3$ & - & - \\
  \midrule
  \rowcolor{Gray!16} SIFT1M & P1 & $0.7537$ & \textcolor{BestGreen}{$12879.1$} & $0.6469$ & $414.0$ & \textcolor{BestGreen}{$0.8387$} & $3177.1$ \\
  \rowcolor{WhiteColr}  & P2 & $0.8734$ & \textcolor{BestGreen}{$6990.6$} & $0.6458$ & $295.0$ & \textcolor{BestGreen}{$0.9003$} & $1683.0$ \\
  \rowcolor{LightCyan}  & P3 & \textcolor{BestGreen}{$0.9997$} & \textcolor{BestGreen}{$611.7$} & $0.6405$ & $91.72$ & - & - \\
  \midrule
  \rowcolor{Gray!16} GloVe & P1 & $0.2670$ & \textcolor{BestGreen}{$8068.4$} & $0.3453$ & $438.5$ & \textcolor{BestGreen}{$0.6543$} & $1927.7$ \\
  \rowcolor{WhiteColr}  & P2 & $0.4845$ & \textcolor{BestGreen}{$5045.1$} & $0.3428$ & $392.5$ & \textcolor{BestGreen}{$0.6543$} & $1940.8$ \\
  \rowcolor{LightCyan}  & P3 & $0.6925$ & \textcolor{BestGreen}{$818.3$} & $0.3379$ & $390.2$ & \textcolor{BestGreen}{$0.8371$} & $212.3$ \\
  \bottomrule[1.5pt]
  \end{tabular}
  \label{tab:ivfpq-qps-recall}
  \vspace{-4mm}
  \end{table}

\subsubsection{Overhead}

\begin{figure}[t]
    \centering
    \includegraphics[width=\columnwidth]{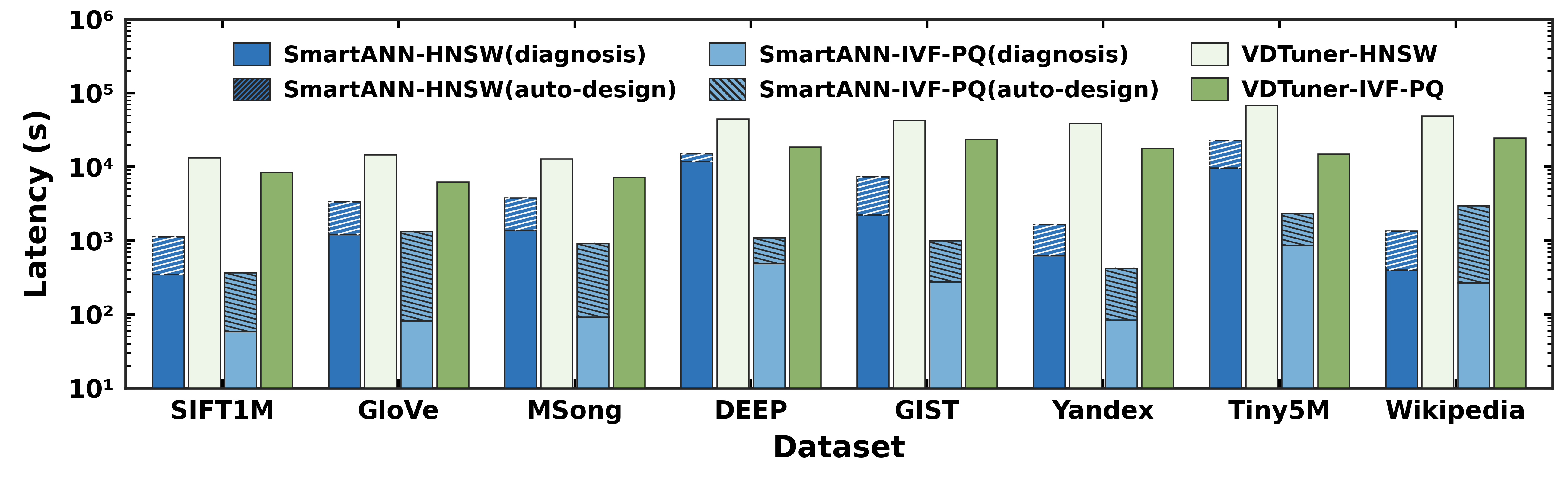}
    \caption{End-to-end tuning time of SmartANN and VDTuner on HNSW and IVF-PQ. Each SmartANN bar separates attribution diagnosis from auto-design.}
    \label{fig:method-cost}
\end{figure}

Figure~\ref{fig:method-cost} compares the end-to-end tuning time of SmartANN and VDTuner. SmartANN time consists of attribution diagnosis and auto-design. Auto-design includes candidate-action training, materialization, and evaluation. VDTuner time covers the complete joint search over construction and query parameters. The vertical axis uses a logarithmic scale.

SmartANN requires less tuning time than VDTuner on every dataset. For HNSW, the per-dataset speedup ranges from $2.9\times$ to $36.4\times$. The aggregate speedup across eight datasets is approximately $5.0\times$. For IVF-PQ, the per-dataset speedup ranges from $4.7\times$ to $42.0\times$. The aggregate speedup is approximately $11.6\times$. VDTuner repeatedly constructs indexes and executes query evaluations over its complete parameter space. SmartANN restricts candidate actions according to the diagnosed bottleneck objects. It therefore avoids ineffective complete-design evaluations and substantially reduces end-to-end tuning cost.

\subsection{Ablation Study}
\label{sec:eval-ablation}

We ablate two key designs of SmartANN. SmartANN versus Random isolates the effect of attribution-guided automated design, while Replacement versus No-Replacement evaluates whether sequential replacement mitigates performance-loss propagation and reveals independent downstream bottlenecks.

\subsubsection{Effect of Attribution-guided Automated Design}

\begin{figure}[t]
    \centering
    \includegraphics[width=\columnwidth]{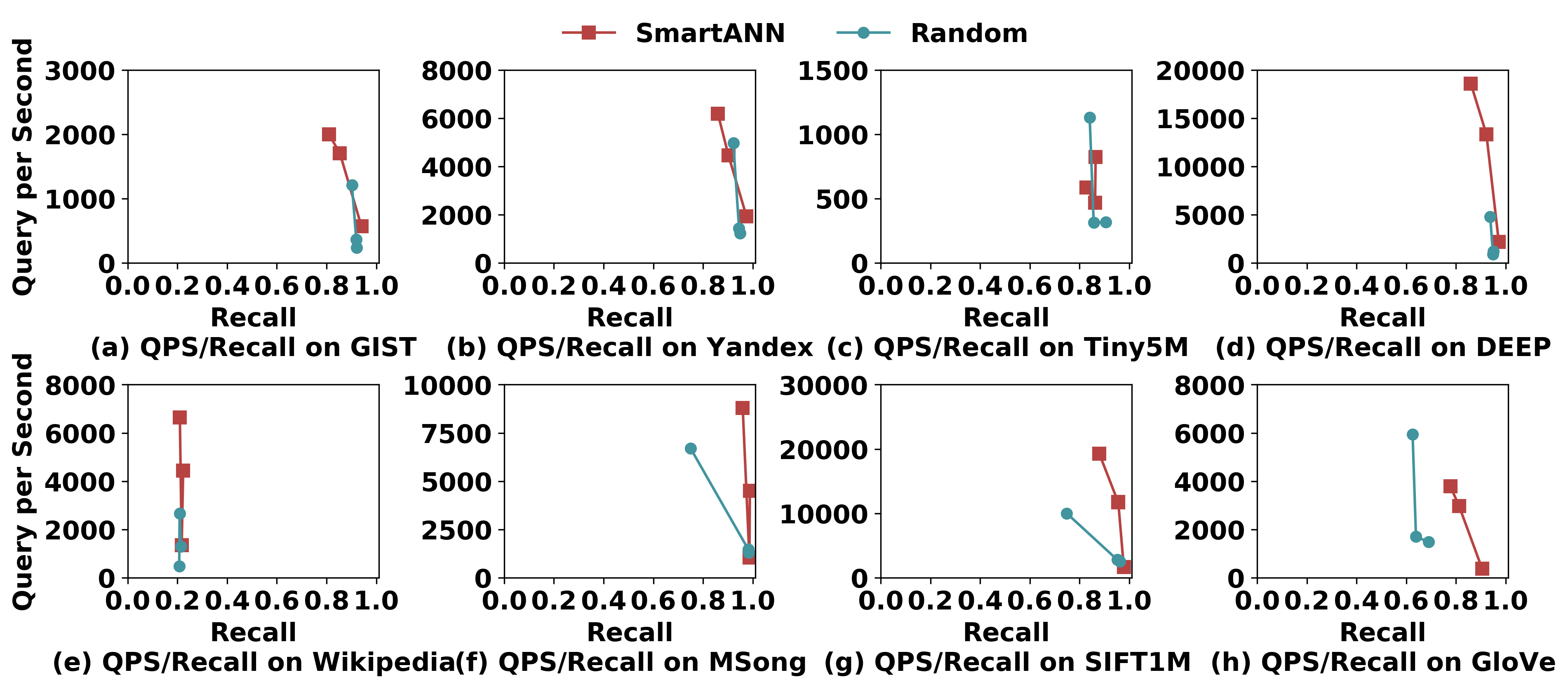}
    \caption{Recall--QPS comparison between attribution-guided SmartANN and Random on HNSW.}
    \label{fig:hnsw-random}
\end{figure}

\begin{figure}[t]
    \centering
    \includegraphics[width=\columnwidth]{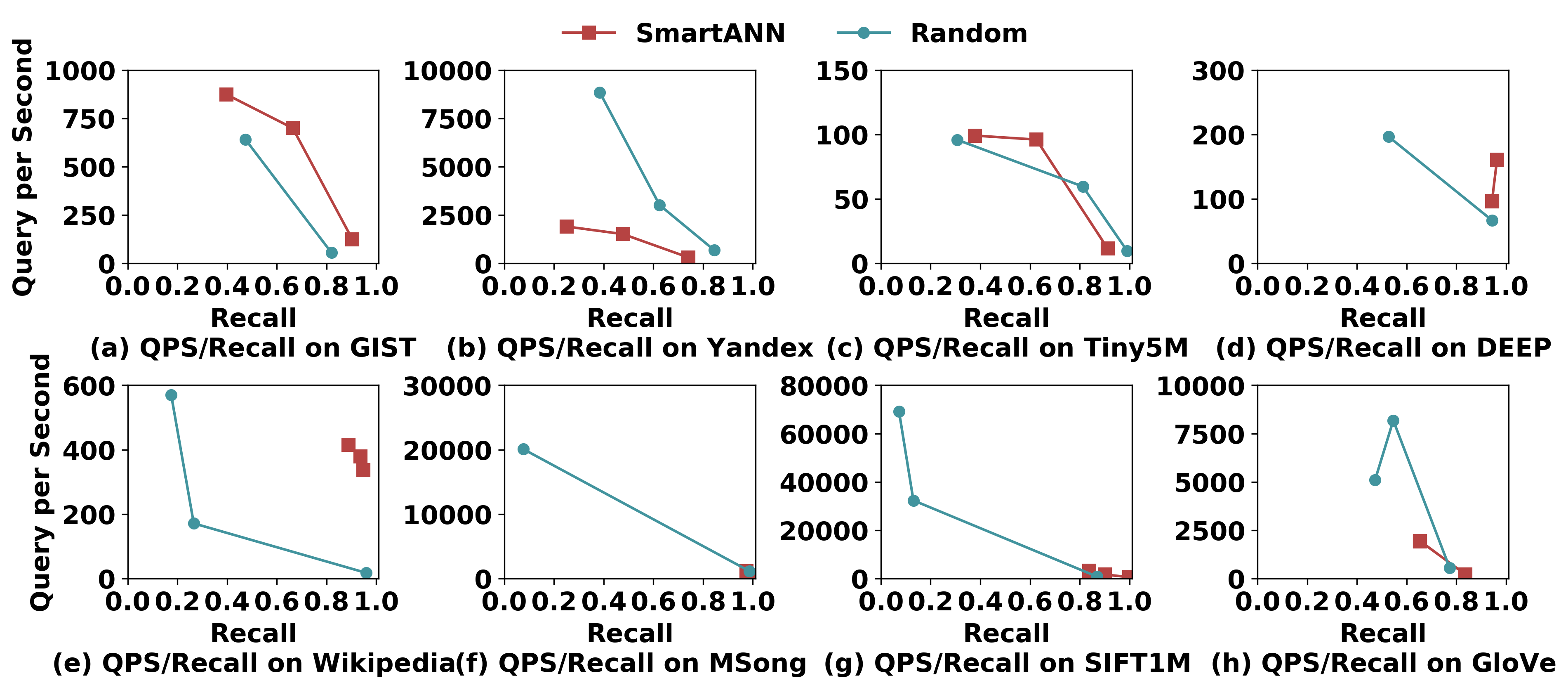}
    \caption{Recall--QPS comparison between attribution-guided SmartANN and Random on IVF-PQ.}
    \label{fig:ivfpq-random}
\end{figure}

Figures~\ref{fig:hnsw-random} and~\ref{fig:ivfpq-random} compare SmartANN with Random on HNSW and IVF-PQ. Both use the same baseline operating points, action library, and resource budgets, but only SmartANN selects actions using the bottleneck-object set and failure causes. Their Recall--QPS differences therefore isolate the benefit of attribution-guided automated design.

SmartANN achieves more consistent Recall--QPS tradeoffs than Random on both HNSW and IVF-PQ. On IVF-PQ GIST, SmartANN reaches 0.9038 Recall and 124.4 QPS, outperforming Random at 0.8195 Recall and 55.6 QPS. Although Random occasionally finds competitive designs, its results are sensitive to action sampling and often gain QPS by sacrificing Recall. These results show that bottleneck attribution more reliably selects actions matching the diagnosed failure causes.

\subsubsection{Effect of Replacement Operator.}
We compare Replacement and No-Replacement using the same baseline execution, object order, diagnostic metrics, and thresholds. After identifying a bottleneck \(O_i\), Replacement substitutes its output with the test oracle or a stronger replacement implementation and re-executes the affected downstream objects. No-Replacement retains the downstream metrics from the original execution. We compare their downstream metrics and resulting bottleneck-object sets.

\noindent\textbf{(1) HNSW.}
Table~\ref{tab:hnsw-replacement} shows both propagated and independent downstream loss. On GloVe, replacing Candidate Exploration at \(O_6\) increases Candidate Recall at \(O_6\) from 0.642 to 0.790 and at \(O_7\) from 0.687 to 0.790. On Yandex-200, the corresponding values increase from 0.898 and 0.884 to 0.956. Because HNSW uses exact vector scoring, the normalized \(O_7\) metrics confirm that its previous deviation was propagated from \(O_6\), rather than generated by an independent bottleneck.

Tiny5M and DEEP contain bottlenecks at both \(O_2\) and \(O_6\). Replacing \(O_2\) changes Candidate Recall at \(O_6\) only from 0.714 to 0.735 on Tiny5M and from 0.864 to 0.860 on DEEP. Replacing \(O_6\) further increases Recall to 0.829 and 0.944, respectively, confirming an independent Candidate Exploration bottleneck.

\noindent\textbf{(2) IVF-PQ.}
Table~\ref{tab:ivfpq-replacement} shows that upstream loss can also mask downstream bottlenecks. On GIST, replacing \(O_2\) reduces true-neighbor-unit coverage at \(O_5\) from 0.906 to 0.885, exposing an Access Localization bottleneck. Replacing \(O_5\) subsequently increases candidate loss at \(O_7\) from 0.242 to 0.338, exposing an independent Candidate Scoring bottleneck. The same cause appears on GloVe and Yandex-200, where improving \(O_5\) coverage to 1.000 increases \(O_7\) candidate loss from 0.184 to 0.421 and from 0.304 to 0.442, respectively. Replacing \(O_7\) reduces these losses to zero.

No-Replacement either attributes propagated upstream loss repeatedly to downstream objects or misses independent bottlenecks masked by upstream loss. Sequential replacement suppresses confirmed performance-loss propagation and re-evaluates downstream objects, enabling SmartANN to identify the complete bottleneck-object set without duplicate attribution.

\begin{table}[t]
\centering
\footnotesize
\setlength{\tabcolsep}{4pt}
\renewcommand{\arraystretch}{1.12}
\caption{HNSW replacement results on diagnosed quality-problem points.
Parentheses show the absolute change relative to the previous row of the \emph{same} dataset and point (current $-$ previous).
The first row of each point, \textit{clean}, unchanged, and missing entries omit $\Delta$.
O2 (hops) lower is better; O6 ($\bar{R}$) and O7 (scored-candidate recall) higher is better.
Undiagnosed O2 is \textit{clean}.
O7 before replacement is selection-batch scored recall; after the O6 oracle it tracks post-O6 $\bar{R}$ (native exact scoring).
O2 replacement is hierarchy preview; O6 replacement is the diagnostic-period O6 oracle.
}
\label{tab:hnsw-replacement}
\resizebox{\linewidth}{!}{%
\begin{tabular}{@{}llccc@{}}
\toprule[1.5pt]
\textbf{Dataset} &
\textbf{replacement} &
\begin{tabular}[c]{@{}c@{}}\textbf{O2}$\downarrow$\end{tabular} &
\begin{tabular}[c]{@{}c@{}}\textbf{O6}$\uparrow$\end{tabular} &
\begin{tabular}[c]{@{}c@{}}\textbf{O7}(not bottleneck)$\uparrow$\end{tabular} \\
\midrule
\multicolumn{5}{l}{\textit{Bottlenecks: O2 + O6}} \\
\rowcolor{Gray!16}
\texttt{DEEP}       & Initial   & $9.89$ & $0.864$ & \na{} \\
\rowcolor{WhiteColr}
                       & O2     & \mvup{9.75}{-0.14} & \mvdn{0.860}{-0.004} & $0.860$ \\
\rowcolor{LightCyan}
                       & O2\&O6 & $9.75$ & \mvup{0.944}{+0.084} & \mvup{0.944}{+0.084} \\
\cmidrule(lr){1-5}
\rowcolor{Gray!16}
\texttt{Tiny5M}     & Initial   & $10.28$ & $0.714$ & \na{} \\
\rowcolor{WhiteColr}
                       & O2     & \mvup{8.31}{-1.97} & \mvup{0.735}{+0.021} & $0.663$ \\
\rowcolor{LightCyan}
                       & O2\&O6 & $8.31$ & \mvup{0.829}{+0.094} & \mvup{0.829}{+0.166} \\
\midrule
\multicolumn{5}{l}{\textit{Bottleneck: O6 only}} \\
\rowcolor{Gray!16}
\texttt{MSong}      & Initial   & \textit{clean} & $0.968$ & 0.967 \\
\rowcolor{WhiteColr}
                       & O6     &  \textit{clean}& \mvup{0.989}{+0.021} & \mvup{0.989}{+0.021} \\
\cmidrule(lr){1-5}
\rowcolor{Gray!16}
\texttt{SIFT1M}       & Initial   & \textit{clean} & $0.940$ & $0.950$ \\
\rowcolor{WhiteColr}
                       & O6     & \textit{clean} & \mvup{0.962}{+0.022} & \mvup{0.962}{+0.012} \\
\cmidrule(lr){1-5}
\rowcolor{Gray!16}
\texttt{GloVe}      & Initial   & \textit{clean} & $0.642$ & $0.687$ \\
\rowcolor{WhiteColr}
                       & O6     & \textit{clean} & \mvup{0.790}{+0.148} & \mvup{0.790}{+0.103} \\
\cmidrule(lr){1-5}
\rowcolor{Gray!16}
\texttt{GIST}       & Initial   & \textit{clean} & $0.761$ & $0.811$ \\
\rowcolor{WhiteColr}
                       & O6     & \textit{clean} & \mvup{0.846}{+0.085} & \mvup{0.846}{+0.035} \\
\cmidrule(lr){1-5}
\rowcolor{Gray!16}
\texttt{Yandex}     & Initial   & \textit{clean} & $0.898$ & $0.884$ \\
\rowcolor{WhiteColr}
                       & O6     & \textit{clean} & \mvup{0.956}{+0.058} & \mvup{0.956}{+0.072} \\
\cmidrule(lr){1-5}
\rowcolor{Gray!16}
\texttt{Wikipedia}  & Initial   & \textit{clean} & $0.275$ & $0.270$ \\
\rowcolor{WhiteColr}
                       & O6     & \textit{clean} & \mvup{0.301}{+0.026} & \mvup{0.301}{+0.031} \\
\bottomrule[1.5pt]
\end{tabular}%
}
\end{table}

\begin{table}[t]
\centering
\footnotesize
\setlength{\tabcolsep}{4pt}
\renewcommand{\arraystretch}{1.12}
\caption{IVFPQ replacement results on eight datasets.
Parentheses show the absolute change relative to the previous row of the \emph{same} dataset (current $-$ previous).
The first row of each dataset and any \textit{clean} or unchanged entry omit $\Delta$.
O2 / O7: lower is better; O5: higher is better.
}
\label{tab:ivfpq-replacement}
\resizebox{\linewidth}{!}{%
\begin{tabular}{@{}ll ccc@{}}
\toprule[1.5pt]
\textbf{Dataset} &
\textbf{replacement} &
\begin{tabular}[c]{@{}c@{}}\textbf{O2}($\downarrow$)\end{tabular} &
\begin{tabular}[c]{@{}c@{}}\textbf{O5}($\uparrow$)\end{tabular} &
\begin{tabular}[c]{@{}c@{}}\textbf{O7}($\downarrow$)\end{tabular} \\
\midrule
\multicolumn{5}{l}{\textit{Bottlenecks: O2 + O5 + O7}} \\
\midrule
\rowcolor{Gray!16}
\texttt{GIST} & Initial      & $5.547$ & $0.906$ & $0.264$ \\
\rowcolor{WhiteColr}
                          & O2        & \mvup{1.813}{-3.734} & \mvdn{0.885}{-0.021} & \mvup{0.242}{-0.022} \\
\rowcolor{LightCyan}
                          & O2\&O5    & $1.813$ & \mvup{1.000}{+0.115} & \mvdn{0.338}{+0.096} \\
\rowcolor{Peach!16}
                          & O2\&O5\&O7 & $1.813$ & $1.000$ & \mvupb{0.000}{-0.338} \\
\midrule
\rowcolor{Gray!16}
\texttt{Yandex}       & Initial      & $5.387$ & $0.800$ & $0.265$ \\
\rowcolor{WhiteColr}
                          & O2        & \mvup{4.158}{-1.229} & \mvdn{0.798}{-0.003} & \mvdn{0.304}{+0.040} \\
\rowcolor{LightCyan}
                          & O2\&O5    & $4.158$ & \mvup{1.000}{+0.203} & \mvdn{0.442}{+0.138} \\
\rowcolor{Peach!16}
                          & O2\&O5\&O7 & $4.158$ & $1.000$ & \mvupb{0.000}{-0.442} \\
\midrule
\rowcolor{Gray!16}
\texttt{Tiny5M}           & Initial      & $5.516$ & $0.924$ & $0.529$ \\
\rowcolor{WhiteColr}
                          & O2        & \mvup{1.854}{-3.662} & \mvup{0.944}{+0.020} & \mvdn{0.531}{+0.006} \\
\rowcolor{LightCyan}
                          & O2\&O5    & $1.854$ & \mvup{1.000}{+0.056} & \mvdn{0.583}{+0.023} \\
\rowcolor{Peach!16}
                          & O2\&O5\&O7 & $1.854$ & $1.000$ & \mvupb{0.000}{-0.583} \\
\midrule
\multicolumn{5}{l}{\textit{Bottlenecks: O5 + O7}} \\
\midrule
\rowcolor{Gray!16}
\texttt{DEEP}   & Initial      & \textit{clean} & $0.939$ & $0.217$ \\
\rowcolor{WhiteColr}
                          & O5        & \textit{clean} & \mvup{1.000}{+0.061} & \mvdn{0.260}{+0.043} \\
\rowcolor{LightCyan}
                          & O5\&O7    & \textit{clean} & $1.000$ & \mvupb{0.000}{-0.260} \\
\midrule
\rowcolor{Gray!16}
\texttt{Wikipedia}    & Initial      & \textit{clean} & $0.967$ & $0.111$ \\
\rowcolor{WhiteColr}
                          & O5        & \textit{clean} & \mvup{1.000}{+0.033} & \mvdn{0.143}{+0.032} \\
\rowcolor{LightCyan}
                          & O5\&O7    & \textit{clean} & $1.000$ & \mvupb{0.000}{-0.143} \\
\midrule
\rowcolor{Gray!16}
\texttt{SIFT1M}             & Initial      & \textit{clean} & $0.821$ & $0.086$ \\
\rowcolor{WhiteColr}
                          & O5        & \textit{clean} & \mvup{1.000}{+0.179} & \mvdn{0.160}{+0.074} \\
\rowcolor{LightCyan}
                          & O5\&O7    & \textit{clean} & $1.000$ & \mvupb{0.000}{-0.160} \\
\midrule
\rowcolor{Gray!16}
\texttt{GloVe}    & Initial      & \textit{clean} & $0.644$ & $0.184$ \\
\rowcolor{WhiteColr}
                          & O5        & \textit{clean} & \mvup{1.000}{+0.356} & \mvdn{0.421}{+0.238} \\
\rowcolor{LightCyan}
                          & O5\&O7    & \textit{clean} & $1.000$ & \mvupb{0.000}{-0.421} \\
\midrule
\multicolumn{5}{l}{\textit{Bottleneck: O7 only}} \\
\midrule
\rowcolor{Gray!16}
\texttt{MSong}            & Initial      & \textit{clean} & \textit{clean} & $0.373$ \\
\rowcolor{WhiteColr}
                          & O7        & \textit{clean} & \textit{clean} & \mvupb{0.000}{-0.373} \\
\bottomrule[1.5pt]
\end{tabular}%
}
\end{table}

\subsection{Key Findings}

\underline{\textbf{ (1) Bottleneck attribution.}}

SmartANN combines object-level diagnostic metrics with sequential replacement to identify the complete bottleneck-object set and failure causes to provide verifiable attribution evidence. The ablation shows that No-Replacement misattributes propagated upstream loss to downstream objects and misses independent bottlenecks masked by upstream performance loss.

\noindent\underline{\textbf{(2) Automated redesign.}}

    Using the diagnosed bottleneck-object set and failure causes to select and compose compatible actions produce effective end-to-end ANN redesigns more consistently than the reproducible Random baseline without attribution evidence.

\noindent\underline{\textbf{(3) Effectiveness and efficiency.}}

    Across eight real-world datasets, SmartANN improves Recall by 0.24--74.20 percentage points or increases QPS by 28.8--256.5\% at comparable Recall. Compared with VDTuner, SmartANN achieves more favorable Recall--QPS tradeoffs across most datasets and a \(2.9\)--\(42.0\times\) speedup in end-to-end latency, demonstrating low diagnosis and auto-design overhead.

\section{Related Work}

ANN search have been extensively studied. ANN indexing and optimization explains how ANN indexes perform and optimize search. ANN benchmarks explain how ANN index performance is measured consistently. Component-level analysis of ANN explains the differences among component designs. SmartANN further answers which mechanisms cause performance bottlenecks under a specific data distribution and query workload and which implementations should be used for targeted replacement.


\subsection{ANN Indexing and Optimization}

IVF-PQ~\cite{jegou2011product} and HNSW~\cite{malkov2018efficient} are state-of-the-practice indexes for space partitioning with quantization and proximity graphs. They are widely used practices in current ANN algorithms. 
SmartANN therefore uses both algorithms for diagnosis, replacement, and auto-design. IVF-PQ limits the search range through data partitioning and reduces storage and distance-computation cost through product quantization~\cite{ivfpq-limit-Jegou-TPAMI}. HNSW constructs a multilayer proximity graph and reduces vector accesses by navigating from upper-layer entries to the base layer~\cite{malkov2018efficient}.

Prior work improves representation, data partitioning, graph structure, query routing, candidate exploration, candidate scoring, and search stopping for these two base indexes. For representation, quantization, and candidate scoring, OPQ~\cite{opqTPAMI14}, ScaNN/AVQ~\cite{guo2020accelerating}, DPQ~\cite{dpqICML20}, RPQ~\cite{rpqICDE24}, and RaBitQ~\cite{rabitqSIGMOD24} improve vector encoding and distance estimation. ADSampling~\cite{adsamplingSIGMOD23}, DADE~\cite{dadePVLDB24}, PEOs~\cite{peosICML24}, and FINGER~\cite{fingerWWW23} reduce unnecessary exact distance computations.

For data partitioning and graph structure, Neural LSH~\cite{lsh2020iclr} and BLISS~\cite{blissKDD22} learn data partitions. RoarGraph~\cite{roargraphVLDB24}, SPANN~\cite{spannNEURIPS21}, and Elpis~\cite{elipisVLDB23} design index structures for cross-modal queries, memory--disk indexes, and partitioned subgraphs. Flash~\cite{flashSIGMOD25}, SymphonyQG\\~\cite{symphonyQGSIGMOD25}, VSAG~\cite{vsagVLDB25}, Starling~\cite{starlingSIGMOD24}, and DEG~\cite{degSIGMOD25} optimize graph construction, graph structure, data layout, or system execution.

For query routing, candidate exploration, and search stopping, Learning to Route~\cite{learningtoroute}, LTR-IVF~\cite{ltr-ivf}, and LEQAT~\cite{leqat} improve query routing. LAET~\cite{laet} and DARTH~\cite{darth} determine where search should stop for different queries. FARGO~\cite{fargo} designs global multi-probing for maximum inner product search. MEVI~\cite{mevi} combines a generative model with a vector index for document retrieval. Steiner-Hardness~\cite{Steiner-Hardness} is originally a query-difficulty measure. 

Existing ANN optimizations usually design complete solutions for predefined problems and validate them through end-to-end performance. They cannot diagnose the bottleneck mechanisms of an ANN index under a new data distribution and query workload or select the corresponding implementations. SmartANN determines the complete bottleneck mechanism set through diagnosis and replacement, then selects corresponding implementations from a pluggable implementation library for targeted replacement.

\subsection{ANN Benchmarks}

Existing ANN benchmarks mainly provide end-to-end evaluation. Their metrics include index construction time, index memory use, retrieval accuracy, and retrieval speed. 
ANN-Benchmarks~\cite{annbench} is one of the most widely used evaluation tools. Big-ANN-Benchmarks~\cite{bigannbench} extends it to billion-scale datasets. 
BigVectorBench~\cite{bvb} targets compound queries in real applications, including range-filtered queries, multi-vector queries, big queries, and multimodal queries. 
It evaluates vector generation quality, cost, and retrieval performance for heterogeneous data. Iceberg~\cite{icrberg} studies downstream tasks in real applications and shows that Recall--performance evaluation based on vector-distance ground truth may not reflect actual task performance.
Filtered-ANN benchmarks extend end-to-end evaluation by comparing attribute and range filtering algorithms, controlling filter selectivity and attribute--vector correlation, and generating workloads by query difficulty~\cite{Zhu2025AnEE,vecbench,hcbgen}.

Existing end-to-end benchmarks can compare the construction cost and retrieval performance of complete ANN indexes. They cannot analyze a specific mechanism at a fine granularity and therefore cannot locate a performance bottleneck under a real data distribution and query workload. SmartANN determines the complete bottleneck mechanism set through diagnosis and replacement. It then selects and replaces corresponding pluggable implementations to improve index performance.

\subsection{Component-Level Analysis of ANN}

Prior research has begun to move from complete-algorithm comparison to component-level analysis. 
Wang et al. divide graph ANN into seven components and compare different implementations of a target component while fixing the others~\cite{wang-survey-vldb2021}. 
Azizi et al. summarize graph ANN into five design categories and perform controlled analysis on selected component~\cite{azizi}s. 
Other studies analyze the effect of predefined stages. 
Gottesb{\"u}ren et al. use exhaustive search within shards and a routing oracle to analyze data partitioning, query routing, and in-shard search~\cite{gottesb}. 
FANNS~\cite{fanns} divides an IVF-PQ query into six stages and analyzes the execution time of each stage. 
Hua et al. divide graph search into two stages and diagnose and repair reachability and local graph structure~\cite{hua-ngfix}. 
Other studies provide fine-grained analyses of pruning and entry selection in filtered ANN~\cite{fannbench}, distance comparison~\cite{discompare}, I/O components in disk ANN~\cite{diskannio}, and graph-index construction~\cite{graphindex}.

Existing component-level analyses first specify the component or problem to study, then compare its implementations or measure its effect. SmartANN does not assume the bottleneck location in advance. It decomposes an ANN index into eight mechanisms and defines the input, output, and diagnostic metrics of each mechanism. The diagnosis--replacement--continued diagnosis process determines the complete bottleneck mechanism set under the current workload, measures the recoverable performance loss of each mechanism, and guides targeted mechanism replacement.

\section{Conclusion}

This paper presents SmartANN, a framework built on the object causal model for ANN bottleneck attribution and automated redesign. SmartANN represents ANN index construction and query execution as eight ordered, replaceable objects. Its sequential \textit{diagnose-and-replace} loop uses test-oracle outputs or stronger replacement implementations to mitigate performance-loss propagation and identify the complete bottleneck-object set and failure causes. SmartANN then automatically selects and composes compatible actions from a pluggable action library to generate optimized end-to-end ANN designs. Extensive experiments on eight real-world datasets across IVF-PQ and HNSW show that SmartANN improves Recall by 0.24--74.20\% or increases QPS by 28.8--256.5\% at comparable Recall, while achieving low diagnosis and auto-design overhead.

\bibliographystyle{plain}
\bibliography{sample}

\end{document}